\documentstyle[twocolumn,aps,prb]{revtex}
\begin{document}
\draft

\title{Acetylene on Si(100) from first principles:
adsorption geometries, equilibrium coverages and thermal decomposition}
\author{Pier Luigi Silvestrelli, Flavio Toigo, and Francesco Ancilotto}
\address{Istituto Nazionale per la Fisica della Materia and
Dipartimento di Fisica ``G. Galilei'',
Universit\`a di Padova, via Marzolo 8, I-35131 Padova, Italy\\}

\date{\today}
\maketitle

\begin{abstract}
Adsorption of acetylene on Si(100) is studied from first principles.
We find that, among a number of possible adsorption configurations,
the lowest-energy structure is a ``bridge'' configuration,
where the C$_2$H$_2$ molecule is bonded to two Si atoms. 
Instead, ``pedestal'' configurations, recently 
proposed as the lowest-energy structures,
are found to be much higher in energy
and, therefore, can represent only metastable adsorption sites.
We have calculated the surface formation energies for two
different saturation coverages, namely 0.5 and 1 monolayer,
both observed in experiments. 
We find that although, in general, the full monolayer coverage is favored,
a narrow range of temperatures exists
in which the 0.5 monolayer coverage is the most stable one, 
where the acetylene molecules are adsorbed in a $2\times 2$ structure.
This result disagrees with the conclusions of a recent study and
represents a possible explanation of apparently 
controversial experimental findings.
The crucial role played by the use of a gradient-corrected 
density functional is discussed.
Finally, we study thermal decomposition of acetylene adsorbed on Si(100)
by means of finite-temperature Molecular Dynamics, and we observe
an unexpected behavior of dehydrogenated acetylene molecules.
\end{abstract}

\vfill
\eject

\noindent
\narrowtext

\section{Introduction}

Adsorption of unsaturated hydrocarbon molecules on Si surfaces
is of great importance to investigate the initial-stage growth
of SiC films on Si substrates.
In particular, there is an increasing interest in the
study of adsorption of two basic hydrocarbon units, namely acetylene 
(C$_2$H$_2$) and ethylene (C$_2$H$_4$),
on Si(100) surfaces.
\cite{Nishijima,Huang,Taylor,Li,Matsui,Terborg,Craig,Zhou,Liu,Imamura,Doren,Dyson,Feng,Dufour,Carmer,Meng,Fisher,Tanida,Sorescu,Xu,Xu-PRB}
Acetylene, due to its reactive triple bond, has a high probability
of interacting with the surface Si dangling bonds, undergoing electronic
rehybridization and sticking to the surface.
Adsorption of acetylene on Si(100) represents a 
promising carbon source for homoepitaxial
growth of cubic SiC because the processing temperature can be
reduced well below 1000 $^ {\circ}$C, thus making it compatible with
the present silicon technologies. 
In fact, at temperatures as high as 600 $^ {\circ}$C, 
acetylene already decomposes,
leading to desorption of hydrogen and the formation of
SiC clusters on the Si(100) surface.
Furthermore, the diffusion of a
small concentration of carbon atoms into the Si substrate
may form a substoichiometric alloy with interesting IR
optical properties.

For the adsorption of acetylene on Si(100), as in the case of
ethylene, most of the experimental and theoretical studies 
support the ``di-$\sigma$'' model,
in which the acetylene molecule is adsorbed across a first-layer 
Si dimer in a {\it bridge} (B) site, with the Si dimers  
preserved (i.e. not cleaved) upon adsorption.
Evidence favoring the B adsorption site comes from 
different experiments,\cite{Nishijima,Huang,Li,Matsui,Terborg}
based on high-resolution electron-energy-loss spectroscopy,
low-energy electron diffraction, X-ray
photoemission spectroscopy, near-edge X-ray-adsorption fine structure,
scanning tunnelling microscopy, photoelectron diffraction, 
and from theoretical calculations.\cite{Craig,Zhou,Liu,Imamura,Doren,Dyson,Meng,Fisher,Tanida,Sorescu}

Recently, however, this scenario has been challenged by Xu 
{\it et al.},\cite{Xu} who studied the adsorption
of ethylene and acetylene on the Si(100)
surface by using a photoelectron holographic imaging technique.
The most surprising result of this study
was the observation of a new preferred adsorption site
for acetylene on Si(100), where the molecule is bonded to
four silicon atoms, in a {\it pedestal} configuration (P) 
between two adjacent Si dimers.
This so-called ``tetra-$\sigma$'' model seems to be consistent 
also with high-resolution
photoemission data.\cite{Xu-PRB}

By analyzing their diffraction images, 
Xu {\it et al.} infer a C-C distance
in adsorbed acetylene of $\sim 1.2$ \AA. Even taking into account the
indeterminacy associated to this estimate 
(which, according to the authors, could be as large as 0.2 \AA), 
this represents another surprising
result. In fact this value is comparable with the
experimental bond length of the triple C-C bond in the free 
acetylene molecule (1.21 \AA), whereas, for 
the $sp^3$-hybridized acetylene adsorbed in the P site,
one would expect a much longer bond, close to the single 
C-C bond of, say, C$_2$H$_6$ (1.54 \AA).

Another somewhat controversial issue concerns the C$_2$H$_2$
equilibrium coverage which is most likely
to be realized on the Si surface. 
Based on experimental measurements, both 0.5 monolayer (ML) 
and 1 ML coverages have been proposed 
as saturation coverages realized, in equilibrium conditions, on the Si(100) 
surface (1 ML means that as many C$_2$H$_2$ molecules are adsorbed 
as Si surface dimers). In particular, 
STM experiments performed at room temperature indicate\cite{Li} 
that the C$_2$H$_2$ molecules
form a local $p(2\times 2)$ or $c(4\times 2)$ structure with a saturation
coverage of 0.5 ML.
Detailed analysis of the STM images reveals lack of adsorption 
on nearest-neighbor sites, thus showing that the C$_2$H$_2$ molecules 
adsorb on alternate dimers:
the two observed surface periodicities depend on whether 
the C$_2$H$_2$ molecules adsorbed on adjacent rows 
are in phase or out of phase. On the other hand
kinetic uptake measurements,\cite{Taylor} performed at lower
temperatures (105 K), 
provide evidence for 1 ML coverage, i.e. each Si dimer site adsorbs
a single C$_2$H$_2$ molecule.
These different experimental observations have been tentatively
explained\cite{Sorescu} by assuming that the early stages of
acetylene chemisorption are kinetically rather than 
thermodynamically driven, i.e.
adsorption proceeds via precursor states,\cite{Taylor}
which although not truly thermodynamically stable states, are separated
by sizeable potential barrier from the stable adsorption 
configurations. 

The process of thermal desorption of acetylene on Si(100) is
also of interest. 
It has been found\cite{Nishijima} that, as the surface temperature is raised 
up to 870 K, the acetylene molecules dissociate, leading 
to the formation of SiH groups and surface carbon. However, 
the details of the dissociation paths and the early stages of
surface carbonization are not known. 
Simulations performed using
ab-initio finite-temperature Molecular Dynamics can provide
valuable information to elucidate such complex phenomena.

In order to clarify the above issues, 
we have performed a full
{\em ab initio} study of acetylene adsorption on Si(100), 
by using the Car-Parrinello approach.\cite{CPMD}

The paper is organized as follows.
In Section II we describe the computational method. 
We report our results for the adsorption geometries,
with particular emphasis on the energetic and structural characterization 
of the B and P configurations, in Section IIIA.
In Section IIIB we address the problem of the theoretical determination 
of the C$_2$H$_2$ equilibrium coverage.
Our calculations of the surface formation energies for different
model structures at 0.5 and 1 ML coverage provide evidence
that the saturation coverage depends on
temperature, largely favoring the 1 ML coverage, but indicating the 
existence, for a particular adsorption configuration,
of a narrow window in temperature (which we 
find only qualitatively in agreement with the experiment), where 
the 0.5 ML coverage may become
favored. We offer this prediction, which disagrees with the conclusions
of Ref. \onlinecite{Tanida}, as a semi-quantitative explanation for the
different experimental observations quoted above.

Finally, in Section IV we present the results of our simulations of
thermal dissociation of C$_2$H$_2$ on Si(100);
this is a very important topic being related with the 
process of growth of SiC films obtained from decomposition of acetylene
chemisorbed on the Si(100) surface.
Our results are encouraging and show some interesting
behavior, including a tendency of the acetylene
molecules to decompose into intact C dimers, which diffuse 
below the Si surface layer and bind themselves to 
form more complex structures.

\section{Method}
Total-energy calculations and Molecular Dynamics
(MD) simulations have
been carried out within the Car-Parrinello approach\cite{CPMD}
in the framework of the density
functional theory, both in the local spin density approximation (LSDA)
and using gradient
corrections in the BLYP implementation.\cite{BLYP}
Gradient-corrected functionals have been adopted in the most
recent theoretical studies of adsorption of organic molecules
on Si(100) because they are typically more accurate
than the LSDA functional in describing chemical processes 
on the Si(100) surface\cite{Sorescu,Nachtigall}.
The calculations have been carried out considering the $\Gamma$-point
only of the Brillouin zone (BZ), and using norm-conserving
pseudopotentials,\cite{Troullier} with $s$ and $p$ nonlocality for
C and Si. Wavefunctions were expanded in plane waves with an
energy cutoff of 50 Ry. We have explicitly checked that, at this value of
the energy cutoff, the structural and binding properties of our system
are well converged.

In order to test the above approximations, we have 
calculated the structural and vibrational properties of the free acetylene
molecule (see Table \ref{C2H2} ) and find an excellent
agreement with the experimental estimates and previous theoretical
calculations.\cite{Borrmann}

The Si(100) surface is modeled with a periodically repeated slab
of 5 Si layers and a vacuum region of 7 \AA~ (tests
have been also carried out with a vacuum region of 10 \AA, without
any significant change in the results).
A monolayer of hydrogen atoms is used to saturate the dangling bonds on the
lower surface of the slab.
We have used a $p(4\times4)$ surface supercell
in the $(2\times 1)$ reconstruction, corresponding to
16 Si atoms/layer (some calculations have been performed using
instead a $(4\times 2)$ reconstruction, see Section IIIB). 
With such a relatively large supercell we expect that a
sampling of the BZ limited to the $\Gamma$ point is adequate.
A sensitive test for the adequacy of the $k$-point sampling is provided by the 
study of the clean Si(100) surface: as discussed below, 
the correct surface structure is obtained with our supercell.

Structural relaxations of the ionic coordinates are performed
using the method of direct inversion in the iterative subspace.\cite{DIIS}
During ionic relaxations and MD simulations
the lowest Si layer and the saturation hydrogens are kept fixed.
We verified that, by starting with the unreconstructed,
clean Si(100) surface, the structural optimization procedure
correctly produces
asymmetric surface dimers, with a dimer bond length and
buckling angle in good agreement with previous, highly
converged {\em ab initio} calculations.\cite{Bertoni}
Acetylene molecules are added on top of the slab and
the system is then fully relaxed towards the minimum energy configuration.
To better explore the complex potential energy surface of this system,
in most of the cases the optimization procedure
was repeated using a simulated-annealing strategy and also starting
from different initial configurations.
The same optimization procedure has been already successfully applied
to the study of adsorption of benzene on Si(100).\cite{PRB}

Among the many possible adsorption configurations\cite{Zhou,Dyson}
for acetylene on Si(100) we have focused our study on the B and
P structures (those involved in the present controversy about the
lowest-energy site), although calculations have been 
performed for other configurations as well. 
Adsorption of acetylene between two adjacent dimer rows is 
not considered in our calculations, since it is known to 
result in structures having much lower binding energies.\cite{Zhou,Dyson}

\section{Results}
\subsection{Lowest-energy adsorption sites}
Four possible configurations of
acetylene adsorbed on Si(100) are shown in Fig. \ref{CONF};
the energetics and the structural properties are 
summarized in Table \ref{data}.
When comparing the total energies of different
adsorption configurations, we consider 
a representative 0.5 ML coverage of acetylene on Si(100),
corresponding to two Si dimers per C$_2$H$_2$ molecule.
This is the coverage 
realized in the experiments by Xu {\it et al.},\cite{Xu-PRB}
and the one which is likely to be prevailing at room temperature.
Note that calculations performed at lower coverage 
(for instance at 0.125 ML coverage, corresponding,
in our supercell, to a single acetylene molecule adsorbed on the slab)
do not alter the conclusions contained in Table \ref{data}, 
as far as the energetic ordering of the structures is
concerned.

We confirm that both the P and B structure are 
possible adsorption configurations, and that
the Si dimers involved in bonding with acetylene are preserved
(in line with the most recent experimental and theoretical
works, see Refs. \onlinecite{Terborg,Tanida} and further references quoted
therein), although the typical buckling of
the (2$\times $1) reconstruction is removed and the Si dimers
become symmetric (this latter result is also in agreement with previous
calculations, see, for instance, Ref. \onlinecite{Meng}). 
According to our data
the P structure, however, lies at much higher energy than the B one.
Interestingly, a lower-energy pedestal configuration (P$\,'$) is
obtained by rotating the P structure by 90$^{\circ}$ with respect 
to the Si surface. In this configuration, the four C-Si bonds are
shorter than in the P one. However, P$\,'$ is also 
found to be less favored than B.
If the B structure is rotated by 90$^{\circ}$, in such a way to
bridge two Si atoms of two adjacent dimers, one obtains
another possible adsorption configuration, B$\,'$, slightly higher in energy
than B.
Our results compare favorably with those of the recent paper 
by Sorescu and Jordan,\cite{Sorescu} where however the 
adsorption energies for the B and B$\,'$ structures at 0.5 ML coverage 
are not reported.
We also find that, at higher coverage, the B$\,'$ structure becomes
favored with respect to the B one (the energy difference 
is $\sim 0.14$ and 0.19 eV/molecule, using the LSDA
and BLYP functional, respectively), again in agreement with
the calculations of Sorescu and Jordan.\cite{Sorescu}

As can be seen in Table \ref{data}, use of BLYP gradient corrections 
instead of LSDA makes
bond lengths about 1-2 \% longer; moreover the
P$\,'$ configuration is much less energetically favored,
and the B and B$\,'$ structures are almost degenerate. 
Note however that the energetic ordering of the four
different structures remains unchanged.

Our results contradict the 
findings of Xu {\it et al.}\cite{Xu} that the P site is the 
{\it stable} adsorption site, and confirm instead the 
conclusions reported in the previous literature.
For the B configuration, which is certainly the favored one at
coverages not higher than 0.5 ML, our structural parameters are in 
excellent agreement both with the experimental estimates
of Matsui {\it et al.}\cite{Matsui} (C-C bond length=$1.36 \pm 0.04$ \AA)
and Terborg {\it et al.}\cite{Terborg} (C-C bond length=$1.36 \pm 0.19$ \AA,
C-Si bond length=$1.83 \pm 0.04$ \AA), and
with the most recent theoretical
calculations.\cite{Imamura,Doren,Dyson,Fisher,Tanida,Sorescu} 

In the B structure our optimized C-C bond length (1.35-1.36 \AA)
is close to that of ethylene (1.33 \AA), suggesting $sp^2$
rehybridization of the C-C bond. $sp^2$ bonding 
is supported by 
Imamura {\it et al.},\cite{Imamura} who reported charge-density plots
where the characteristic $\pi$-bonds are observed, and by
Matsui {\it et al.},\cite{Matsui}
who observe $\pi^*$ features in near edge X-ray adsorption
fine structure spectra.
Our computed vibrational frequency of the C-C stretching mode
is 1440 cm$^{-1}$, which compares favorably both with the experimental
value of Huang {\it et al.}\cite{Huang} (estimated to be in the 
range 1450-1500 cm$^{-1}$ ) and with
the previous ab-initio estimate by Imamura {\it et al.}\cite{Imamura}
(1479 cm$^{-1}$).   

Note that, as expected, the C-C bond length in the P and P$\,'$ structures is  
significantly larger 
than in the B one, and is not consistent
with the value, $1.2\pm 0.2$ \AA, observed by Xu {\it et al.}\cite{Xu}.
This leads to the conclusion that, either the experimental 
indeterminacy associated to
the C-C bond length is largely underestimated, or the structure
observed by Xu {\it et al.} is not a pedestal one.

For the B structure, again at 0.5 ML coverage, our computed 
adsorption energy (78 kcal/mol
using LSDA, 65 kcal/mol with BLYP) is in good agreement with
previous theoretical estimates\cite{Imamura,Doren,Dyson,Fisher,Sorescu}
(ranging from 55 to 77 kcal/mol).
The substantial energy difference between the bridge (B, B$\,'$) structures
and the pedestal (P, P$\,'$) ones strongly supports
the conclusion that the bridge adsorption sites are the
{\it stable} configuration for acetylene on Si(100), while the
pedestal structures can represent only {\it metastable} 
adsorption sites.

Other possible adsorption structures exist, besides 
those shown in Fig. \ref{CONF} (see also Ref. \onlinecite{Sorescu}).
For instance, we found a ``diagonal-bridge'' (DB)
configuration in which the acetylene molecule forms
a cross-bridge structure with two Si atoms belonging to two
neighbour Si dimers of the same row.
This configuration can occur as one considers 
reconstructions of the Si surface involving alternating buckled Si dimers,
such as the $p(2\times 2)$ and the $c(4\times 2)$.  
However, according both to our calculations and to those of
Sorescu and Jordan,\cite{Sorescu} the DB structure 
is even energetically less favored than the P$\,'$ one.

\subsection{Saturation Coverage}
Having assessed the quality of our approach
by comparing our structural and energetical data with
previous theoretical and experimental studies, we are now
in the position to deal with the issue of saturation coverage,
where a high accuracy is essential to get meaningful 
information.

STM measurements\cite{Li} of C$_2$H$_2$ on Si(100), 
performed at room temperature, 
indicate a saturation coverage of 0.5 ML (one acetylene molecule adsorbed every
two Si surface dimers).
However, other experimental measurements\cite{Taylor} suggest
instead a saturation coverage of 1.0 ML. 
We show in the following that, although the 1.0 ML coverage seems to
be generally preferred, in principle both coverages can be 
realized depending on the temperature of C$_2$H$_2$ deposition and on the
adsorption configuration.

The 0.5 ML coverage can be realized in a number of ways, even assuming
that each of the acetylene molecule is adsorbed in the lowest 
energy structure, i.e. the B structure (see Fig. \ref{ConfTanida}).
In particular, as shown in the 
following, the relative energy ordering is influenced by the structure
of the Si dimers which are not saturated by C$_2$H$_2$ molecules.
These may be characterized by a different relative buckling orientation.
Here we adopt the same notation used in Ref. \onlinecite{Tanida}
to identify these different structures.
The full monolayer (1 ML) coverage, on the contrary, 
is compatible only with a single configuration if the acetylene
molecules are adsorbed in the B structure.

Using first-principles calculations Tanida {\it et al.}\cite{Tanida}
predict, for the most stable 0.5 ML coverage structure, one (structure (a)
in Fig. \ref{ConfTanida}) where alternated
dimer rows are fully covered with C$_2$H$_2$ molecules.
However such structure is not compatible with
the STM observations of Ref. \onlinecite{Li}, performed
at room temperature.
According to these observations,
at 0.5 ML coverage the acetylene molecules are instead adsorbed 
on alternate dimers in the same rows, and on adjacent rows
they can be in phase or out of phase.
Note that instead the (c)-(f) configurations, which are characterized by 
$(2\times 2)$ or $(4\times 2)$ periodicity, are all
compatible with the STM observations of Ref. \onlinecite{Li}.

Using the BLYP functional we confirm (see Table \ref{data})
the result of Ref. \onlinecite{Tanida} that the (a) structure is
at the lowest energy.
However, at variance with Ref. \onlinecite{Tanida},
we find that the (a), (c), and (e) structures are 
almost degenerate in energy. 
This discrepancy could be due to the fact that Tanida 
{\it et al.}\cite{Tanida} use a smaller cell than ours
and possibly to the different gradient-corrected functional they adopt. 
Note that in this case the effect of the gradient corrections
is not a simple quantitative correction to the LSDA results;
in fact using the BLYP functional the energetic ordering of the
different configurations is completely changed (see Table \ref{data}).

At thermal equilibrium the optimal surface structure can be 
determined by computing the grand canonical potential.\cite{Qian}
We show in Fig. \ref{COVERAGE1}, for the structures (a) and (e) of Fig.
\ref{ConfTanida}, the behavior of $\Delta\Omega$,
that is the difference between the grand canonical potential of a 
given structure and that of the clean Si(100) surface.
$\Delta\Omega$ is defined\cite{Tanida} as
\begin{equation}
\Delta\Omega = \Delta E - N_{_{{\rm C}_2{\rm H}_2}}\mu_{_{{\rm C}_2{\rm H}_2}}\;, 
\end{equation}
where $\Delta E$ is the difference between the total energy of the
considered structure and that of the clean Si(100) surface (using the
lowest-energy surface reconstruction which is compatible with the
given configuration, that is $(4\times 2)$ for configuration (a)
and $(2\times 1)$ for configuration (e), respectively),
and $N_{_{{\rm C}_2{\rm H}_2}}$ is the number of the C$_2$H$_2$ molecules;
following Ref. \onlinecite{Tanida} the zero of the chemical potential
$\mu_{_{{\rm C}_2{\rm H}_2}}$ is assumed to be 1/3 of the energy of the
isolated benzene (C$_6$H$_6$) molecule in vacuum. 
Calculations were performed using the BLYP gradient-corrected functional.
As can be seen, for structure (a) the preferred saturation coverage 
appears to be 1.0 ML, in agreement with the findings of Tanida 
{\it et al.};\cite{Tanida} and the same is true for the structures (b), 
(c), (d), and (f) (the corresponding $\Delta\Omega 
(\mu_{_{{\rm C}_2{\rm H}_2}})$ curves are not shown because they
closely resemble those of structure (a)). 
However, in the configuration (e), there is instead
a range of values of the chemical potential for which
the 0.5 ML saturation coverage becomes favored.  
Note that, considering the adsorption structure B$\,'$ in place of
B, the 1.0 ML saturation coverage would be always favored,
since this structure is the preferred one at high coverage
(see previous discussion).

Configuration (e), with a $2\times 2$ periodicity, 
is compatible with STM measurements at room
temperature and, according to our calculations (see Table \ref{data}),
is also one of the lowest-energy configurations at zero temperature,
being almost degenerate with configuration (a); 
since it is the only structure for which the 0.5 ML coverage could be favored 
instead of the 1.0 ML coverage, we have tried to estimate 
in which range of temperatures this particular configuration is stable.
To this aim we have followed the procedure described in Ref. \onlinecite{rosa}.
The surface formation energy, whose minimum determines 
the equilibrium state of the surface as a function of composition,
is defined as $\Omega /A$, $A$ being the surface area and

\begin{equation}
\Omega (T) =F-\sum _i N_i \mu _i 
\end{equation}
is the grand canonical potential (we have omitted a $PV$ term which, 
for the pressures considered here, is negligible). 
$F$ is the free energy of the system, $N_i$ is the number of particles
of type $i$, and $\mu _i$ denote the chemical potentials of 
the various components.
If one assumes that the Si(100) surface is in equilibrium with bulk Si 
(this is the natural reservoir for Si atoms, due to the
presence of steps, terraces and other surface defects),
then for the bare Si(100) surface one has:

\begin{equation}
\Omega_{\rm bare} = E[{\rm Si}(100)] -N_{\rm Si}E_{\rm bulk}[{\rm Si}]\;, 
\end{equation}
where the first term is the total energy of the slab and the second term
is the energy per atom of bulk Si multiplied
by the total number of Si atoms of the cell.
In the above expression 
the temperature dependence of the free energy
of the slab and that of the bulk reservoir have been neglected,
as usually done in this kind of calculations.\cite{rosa}
The formation energy of the acetylene-covered surface 
can be estimated, following the same reasoning, by computing

\begin{equation}
\Omega (T) = 
E[{\rm Si(100)}{\rm :}{\rm C}_2{\rm H}_2] -N_{\rm Si}E_{\rm bulk}[{\rm Si}]
-N_{_{{\rm C}_2{\rm H}_2}}\mu_{_{{\rm C}_2{\rm H}_2}}(T) + F_{\rm ads}(T)\;,
\end{equation}

where $F_{\rm ads}$ is the free-energy contribution due to the vibrations of
the adsorbed C$_2$H$_2$ molecules.
In order to estimate the temperature dependence of $\mu_{_{{\rm C}_2{\rm H}_2}}$
we assume, quite reasonably, that the surface is in equilibrium
with a reservoir of acetylene molecules in the gas phase.
Therefore, using elementary statistical mechanics,\cite{Kittel}

\begin{equation}
\mu_{_{{\rm C}_2{\rm H}_2}} (T)=
E_{_{{\rm C}_2{\rm H}_2}}
+kT{\rm ln}(PV_Q/kT)-kT{\rm ln}Z_{\rm rot}-kT{\rm ln}Z_{\rm vib}\;.
\end{equation}

Here $E_{{\rm C}_2{\rm H}_2}$ is the total energy of the isolated 
acetylene molecule, 
$P$ is the pressure, and $V_Q=(h^2/2\pi m kT)^{3/2}$ is the quantum volume.
The rotational and vibrational contribution to the
chemical potential are calculated using the experimental
values of the vibrational frequencies for the free 
acetylene molecule.\cite{Zhou}
We estimate the pressure $P$ from realistic 
growth conditions of C$_2$H$_2$ films on Si(100):
experimentally\cite{Nishijima,Li} the saturation coverage is realized
under an acetylene exposure of $\sim 2$ L, which corresponds  
to $P\sim 2\times 10^{ -6}$ Torr.
The vibrational contribution 
$F_{\rm ads}(T)=\sum_i [\hbar\omega_i /2+kT{\rm ln}
(1-{\rm exp}(-\hbar\omega_i/kT))]$ is estimated
from the vibrational frequencies calculated in Ref. \onlinecite{Imamura} 
(the frequencies of the hindered modes,\cite{Nishijima} T$_x$, 
T$_y$ and R$_z$, which are not included among the mode frequencies 
reported in Ref. \onlinecite{Imamura}, have been estimated here 
by using a ``frozen-phonon'' approach).

Our calculations show that, in the temperature range from
410 to 460 K, the adsorption configuration (e) at 0.5 ML coverage
is energetically favored with respect to that corresponding to 1 ML
coverage. Since this temperature interval is not far from room temperature
(given the approximations involved in our derivation 
the accuracy of our temperature estimate is of course limited) we conclude that
configuration (e) could be realized as a stable adsorption configuration
at room temperature.  
This result, which disagrees with the conclusions of Tanida {\it et al.}
\cite{Tanida}
(according to Ref. \onlinecite{Tanida} the 0.5 ML coverage is never favored),
is in qualitative agreement with the
experimental observation\cite{Taylor} that full monolayer 
coverage is realized at $T=105$ K,
while the 0.5 coverage is observed in STM experiments,\cite{Li}
performed at room temperature.
Of course this does not rule out the possibility that
the existence of precursor states\cite{Taylor,Sorescu} could
play a crucial role in destabilizing the surface species for
coverages greater than 0.5 ML; however it offers an 
alternative/complementary explanation of the experimental findings.  

\subsection{Thermal decomposition of C$_2$H$_2$ on Si(100)}
As a result of the high desorption activation energy of C$_2$H$_2$ on
Si(100), 
adsorbed acetylene molecules tend to be 
retained on the surface up to relatively high temperatures.
According to the EELS and LEED measurements of Nishijima 
{\it et al.},\cite{Nishijima} upon heating the
C$_2$H$_2$-exposed Si(100) surface from 80 up to 750 K,
the C-H bond scission occurs, with a gradual recombination of
the dissociated H atoms onto the Si(100) surface; the C-C bond scission
takes place by further heating the sample at 870 K, while by heating 
up to 930 K the surface hydrogen is probably completely removed by
H$_2$ desorption; similar qualitative results have been
obtained in theoretical calculations by Zhou et al.\cite{Zhou}

The proposed\cite{Zhou} decomposition path is therefore:
C$_2$H$_2$ $\rightarrow$ C$_2$H$_x$ ($x$=0,1) $\rightarrow$ C
$\rightarrow$ $\beta$-SiC;
the activation barrier for the dissociation of a single
H atom is estimated\cite{Zhou} to be no less than 0.8 eV,
while that for the the dissociation of the remaining
C$_2$ species is about 1.3 eV.
This indicates that acetylene favors dehydrogenation
to C$_2$H$_x$ ($x$=0,1) species first; then the remaining C$_2$ dimers 
further decompose into C species to form $\beta$-SiC films.
However it is not at all clear\cite{Li,Stedile} if, after acetylene 
dehydrogenation, all the H atoms desorb as a gas of H$_2$ molecules or
if instead some of them actually form Si-H bonds. 
Also the weakening of the Si dimer bonds, as a consequence of 
heating the sample, could be relevant: according to Imamura et al.,
\cite{Imamura} although the dimerized structure is more
stable than the dimer-cleaved one, the latter structure
may be taken instantaneously just after the adsorption, and
the Si-Si bond-breaking and rebonding oscillations could occur
at relatively high temperatures.
Finally, in view of the SiC-growth characterization, it would
be very interesting to clarify how, after acetylene
dissociation, carbon atoms diffuse into the bulk of the Si crystal. 

We have performed Car-Parrinello MD simulations
at finite temperature in order to investigate the
dissociation of acetylene on Si(100).
The ionic degrees of freedom have been integrated using a time step of 5 a.u.
($\sim 0.12$ fs) and typical MD runs were 3-4 ps long.
We start with a configuration where the C$_2$H$_2$ molecules
are adsorbed on the Si(100)$(2\times 1)$ surface in the 
B structure at 0.5 ML coverage.
The initial geometry is optimized using the method described in 
Section II.
We gradually heat this structure by simply rescaling the 
ionic velocities.
Obviously we are forced to use heating rates much higher than
in actual experiments in order to let the system undergo
some kind of transformations; this however 
would lead to melting of the whole
sample prior to C$_2$H$_2$ desorption/decomposition. In order to overcome this
unphysical effect, we have adopted a simple $z$-dependent heating scheme, 
in which the temperature decreases as one moves towards the 
interior of the slab. In this way the
adsorbed acetylene molecules and the uppermost Si atoms are kept at
higher temperatures than the remaining ions, 
thus avoiding melting of the whole slab.
Although both in the simulations and in actual experiments 
the details of the heating procedure may significantly affect 
the final system configuration, we believe that our scheme is able to
reproduce, at least at a qualitative level, the basic features of
the desorption process.

The main results of our MD simulations can be summarized as follows
(given the heating procedure followed, our ``simulation temperature''
is not particularly meaningful for comparison with the experimental one):
1) by heating the sample the first relevant event is the
dissociation of some C-H bonds (typically one H atom
is detached from every molecule);
2) then C-Si bonds are continously broken and reformed
leading to diffusion of the acetylene molecules on the
Si surface;
3) by further increase in the temperature, some
C$_2$H$_2$ molecules loose their H atoms,  
remaining on the surface as C-C dimers; 
4) these dimers remain intact even at very
high temperatures;
5) the dissociated H atoms prefer to form new bonds
with Si atoms rather than forming H$_2$ molecules, although
in our simulation this tendency could be enhanced by the
finite size of the vacuum region;
6) the C-C dimers eventually form complex sub-surface
structures, just below the first Si layer, 
such as a 5-atom ring composed by one Si atom
and 4 C atoms.

Some of the above results are consistent with experimental observations.
In fact, in the early stages of annealing, a strong
Si-H stretching mode is visible in the HREELS spectrum,\cite{Li,Stedile}
thus supporting the formation of Si-H bonds,
while the peak due to C-C stretching broadens and
develops a long tail; this latter result could indicate that
breaking of C-C dimers could be only partial or/and
much slower than H dissociation.
The surprising stability of C-C dimers has been confirmed by using
different heating rates and also adopting the LSDA
functional instead of BLYP. This remarkable effect is in contrast
with previously proposed desorption mechanisms which
just assume a simple C-C dimer thermal breaking
(see for instance Ref. \onlinecite{Nishijima}).
Note that the same result has been obtained by
Cicero {\it et al.}\cite{Cicero} from simulations
in which they studied the heating of carbon dimers
adsorbed on Si(100).

This however poses the problem of how SiC (whose formation is believed to
require breaking of C-C dimers) is actually produced.
A possible explanation could be the presence, in real experiments, 
of a gas of silanes (such as Si$_2$H$_6$)
which could play a crucial role\cite{Fuyuki} in the SiC growth: in fact
adsorbed Si atoms can react with the C$_2$H$_2$ molecules, and
C-C dimer breaking could actually occur before acetylene adsorption.

\section{Conclusions}
Adsorption of acetylene on Si(100) has been investigated
using {\it ab initio} simulations. The structural and
energetic properties of possible adsorption configurations
have been studied.
Although the P structure, recently suggested by
Xu {\it et al.} as the lowest-energy configuration, 
is found to be stable, our results show that the bridge structures
(B or B$\,'$, depending on the coverage)
are energetically favored, in agreement with 
most of the previous studies.
The results reported by Xu {\it et al.} could be influenced by the
presence of defects, such as steps, which could make the actual
Si surface used in their experiment substantially different from the
ideal Si(100) surface assumed (see Fig. 2 of Ref. \onlinecite{Xu}) in the
interpretation of the experimental data.

We have then addressed the issue of the saturation coverage and
its dependence on the temperature, and we find that, although
the full monolayer coverage is generally preferred, there is
a temperature window, not far from room temperature, in which a 
particular adsorption structure at 0.5 ML coverage is preferred.
This could explain the results of STM observations performed at
room temperature.
The effects of including gradient corrections have been also
discussed: our calculations show that these effects are 
particularly relevant for the determination of the optimal
adsorption configuration at 0.5 ML coverage, where even 
fine structural details,   
such as the relative orientation of the buckled Si-dimers, may 
be important for the correct determination of the
surface formation energies.

Finally we have reported the results of our MD simulations
of the thermal decomposition process of acetylene molecules on Si(100);
in particular we observe a surprising stability of the (dehydrogenated) 
C dimers and their tendency to diffuse below the
first Si layer and to coalesce into more complex C fragments. 

\acknowledgments
We acknowledge financial support from INFM, through 
the PRA ``1MESS'', and allocation of computer resources
from INFM ``Progetto Calcolo Parallelo''.
We thank M. Boero, A. Catellani, G. Cicero, R. Di Felice, and G. Onida
for useful discussions.

\vfill
\eject

\begin{figure}
\caption{Stable structures of acetylene adsorbed on Si(100),
in the {\hbox {$(2\times 1)$}} reconstruction:
B=``bridge'', B$\,'$=``rotated bridge'',
P=``pedestal'', P$\,'$=``rotated pedestal''.
For clarity only the four Si atoms of two dimers and
four belonging to the second layer are shown.}
\label{CONF}
\end{figure}

\begin{figure}
\caption{Different possible structures of acetylene adsorbed on Si(100)
in the B configuration, at 0.5 ML coverage: up atoms of Si dimers
are indicated by solid circles, down atoms of Si dimers (and those
belonging to symmetrized Si dimers) by open circles,
and adsorbed acetylene molecules by solid bars.}
\label{ConfTanida}
\end{figure}

\begin{figure}
\caption{Grand canonical potential for the Si(100)-C$_2$H$_2$ surface
at zero temperature, as a function of the chemical
potential $\mu_{_{{\rm C}_2{\rm H}_2}}$, for the structures 
(a) and (e) shown in Fig. 2,
at 0.5 ML coverage (dashed line); data are compared with those
obtained at 1.0 ML coverage (solid line). The values are relative
to the grand canonical potential of the clean Si(100) surface 
(horizontal line).}
\label{COVERAGE1}
\end{figure}

\vfill
\eject

\begin{table}
\caption{Bond length distances and vibrational frequencies of the
free acetylene molecule. 
For comparison, besides the experimental values, we also report the 
LSDA results from Ref. 26
(in parenthesis).}
\begin{tabular}{ccrr}
&LSDA&BLYP&expt.\\ \tableline
C-C (\AA)&1.201 (1.200)&1.206&1.203\tablenote{Ref. 27} \\
C-H (\AA)&1.075 (1.073)&1.066&1.061$^{\rm a}$ \\
C-H stretching (cm$^{-1}$)&3304&3337&3374\tablenote{Ref. 28} \\ 
C-C stretching (cm$^{-1}$)&1975&1960&1974$^{\rm b}$ \\ 
C-H bending (cm$^{-1}$)&739&743&729$^{\rm b}$ \\ 
\end{tabular}
\label{C2H2}
\end{table}

\begin{table}
\caption{Total energy per adsorbed molecule (relative to the
B structure), $E$, and structural parameters
for acetylene on Si(100), at the saturation coverage
of 0.5 ML. ``Si'' indicates the silicon atoms involved in bonding
with acetylene. Data have been computed using the gradient-corrected 
BLYP (Ref. 23) functional. Values obtained using the LSDA functional
are instead reported in parenthesis. 
When slightly different distances are found the average
value is given.
See text for the definitions of the different structures.} 
\begin{tabular}{lcccc}
Configuration&$E$(eV)&$d_{\rm C-C}$(\AA)&$d_{\rm C-Si}$(\AA)&$d_{\rm Si-Si}$(\AA)\\ \tableline
B&0.00 (0.00)&1.36 (1.35)&1.91 (1.88)&2.37 (2.34)\\
B$\,'$&0.05 (0.27)&1.37 (1.36)&1.92 (1.89)&2.43 (2.36)\\
P&1.98 (1.94)&1.58 (1.54)&2.07 (2.02)&2.37 (2.33)\\
P$\,'$&1.28 (0.53)&1.57 (1.53)&2.01 (1.96)&2.32 (2.28)\\
\end{tabular}
\label{data}
\end{table}

\begin{table}
\caption{Total energy (in meV) per adsorbed molecule (relative to the
(a) structure) for different structures of acetylene adsorbed on Si(100)
in the B configuration (see Fig. 2), at 0.5 ML coverage and 
zero temperature, obtained using the
LSDA and BLYP functional, respectively.}
\begin{tabular}{crr}
Configuration&LSDA&BLYP\\ \tableline
(a)&0&0\\
(b)&-21&89\\
(c)&-35&2\\
(d)&-22&18\\
(e)&-31&3\\
(f)&-26&17\\
\end{tabular}
\label{energy}
\end{table}

\vfill
\eject

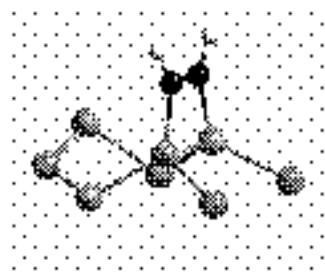
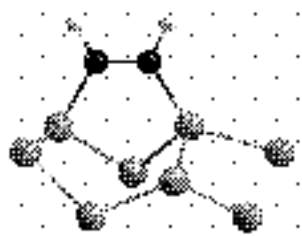
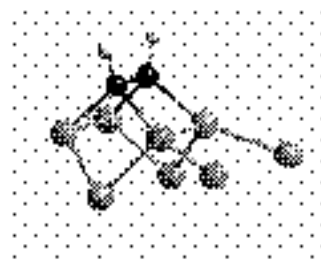
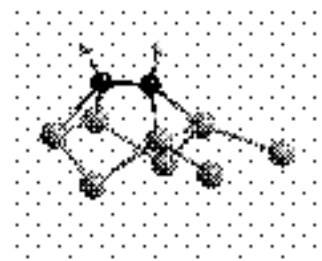
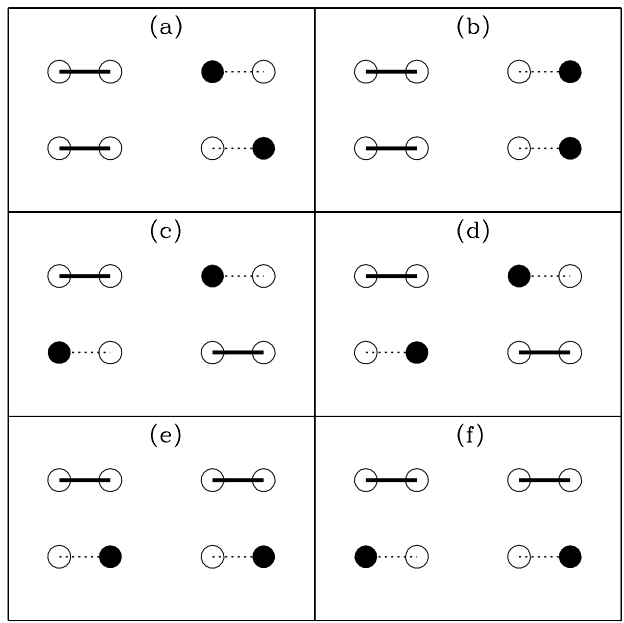
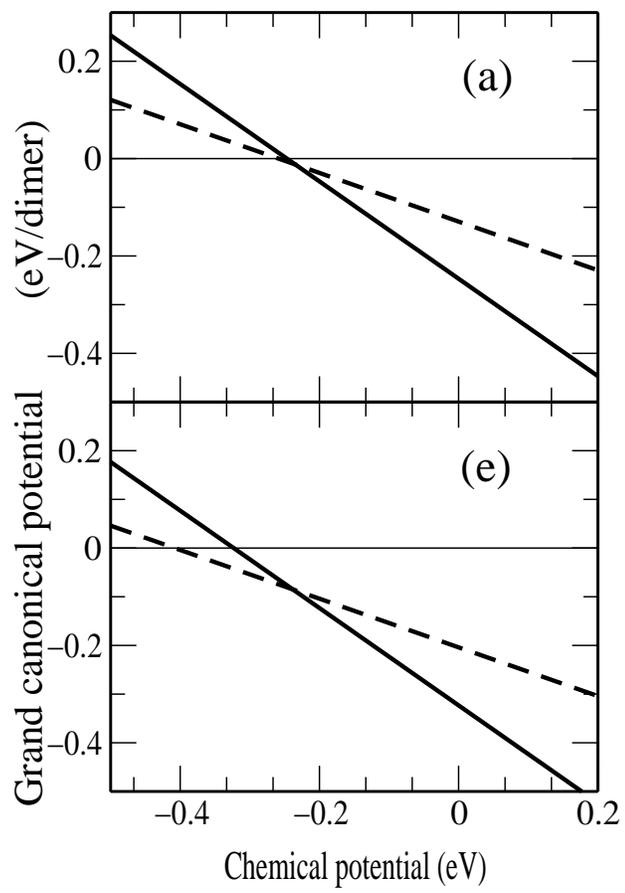
\end{document}